# FABRICATION PROCESS OF ROUNDED DAMPED DETUNED STRUCTURE

N. Hitomi, Y. Funahashi, Y. Higashi, T. Higo, S. Koike, T. Suzuki,
K. Takata, T. Takatomi, T. Toge, Y. Watanabe
KEK, High Energy Accelerator Research Organization, Oho 1-1, Tsukuba, Ibaraki, 305-0801, Japan

*Abstract*

Following the successful design and fabrication of Damped Detuned Structures (DDS), the JLC/NLC linear collider project advanced to Rounded Damped Detuned Structures (RDDS) with curved cross section of the cavity shape for increased shunt impedance. Various advanced techniques for fabricating RDDS1 disks comparing to those for DDS were established to satisfy the dimension accuracy of ±1µm over the entire surface made by ultra-precision turning. These disks were assembled with almost the same stacking and bonding jigs and processes as those of DDS3 assembly. In consequence, the assembly showed little disk-to-disk misalignment within 1µm before and after the process. Though, it had 200µm smooth bowing, which was subsequently corrected as DDS3, and flares at both ends.

## 1 INTRODUCTION

We have been developing the fabrication technologies containing both diamond turning and bonding for Detuned Structure (DS) and DDS. As for the cutting process, work fixture, tool, cutting condition, inspection systems and so on were developed. However, since these technologies were optimised for cutting flat and cylindrical surfaces in DS and DDS, it became necessary to re-study on the procedures for cutting and quality-assuring the curved surface profile of RDDS with good accuracy [1]. As for the stacking disks and bonding them, we already have some jigs for stacking and two stage bonding, carrying out the diffusion bonding following the pre-bonding with some process parameters [2]. We improved some of the processes and succeeded in substantially reducing the magnitude of "bookshelving" stack alignment errors, although we have still fairly large bowing and end flare problems [3].

In the present paper, we describe the main technologies developed for the ultra-precision cutting, precise stacking and the two stage bonding of disks to meet the requirements of various accuracies.

## 2 DIMENSION REQUIREMENTS

Several DS and DDS were designed and fabricated to study the feasibility of realizing the requirements in X-band disk loaded structure. With successful results from these early studies, JLC/NLC linear collider project advanced to rounding the cavity shape to obtain better shunt impedance and termed it RDDS. The cross section of the RDDS is as shown in Fig. 1. The shape of the cross section of a cavity consists of combination of circular and straight lines. The diameter of beam hole (2a) and that of equator (2b) in each disk are different by several microns. The accuracy of cavity contour dimension is specified as within 2 µm. The tolerance for the width of 3mm slit is set to be ±15 µm, since the cell frequency is relatively insensitive to their errors.

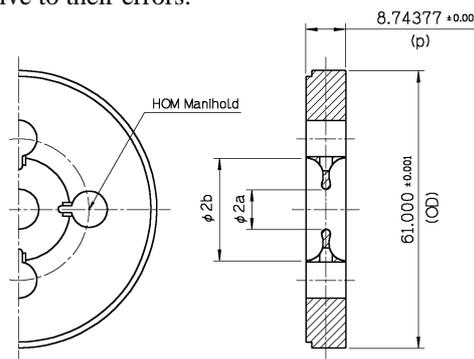

Figure 1: RDDS disk

The surface roughness of the diamond turned surfaces are equal to or less than 50nm. The concentricity between input- and output- side curves of the cavity is specified to be within 0.5µm to obtain a better connection between surfaces in both sides of a disk.

## 3 DIAMOND TURNING

### 3.1 Machine and operation

The temperature of the cutting room and the ultra precision lathe is controlled with an air flow at 20±0.3 °C. The lathe is installed inside of a vinyl cloth hut in the room to keep the temperature and also to prevent dust. This lathe has two axes, namely, one is the Z-axis, the rotating axis of the spindle with a chucking jig to hold the work disk and another the X-axis the tool rest to fix the diamond tool. Both axes are controlled by Computer Numerical Control (CNC) servo-system with 10 nm step glass scales. To obtain the best condition of whole cutting system, the lathe is operated for two hours long before initiating the actual cutting operation.

As the milling process of HOM holes and slits were already finished in the roughing process by a vender with good accuracy, we used this lathe alone as mid- and final-turning process. We cut the surface with rather quick operation in the mid-turning and left about 10μm stock. As a final turning process, we peeled several times. The final cut is the 1μm-deep peel-cut with 3mm/min feed especially at the torus shape edge of beam hole. The dimension data in CNC is corrected by the temperature term modification according to the temperature deviation, that is over than ±0.2 ºC compared with the surface temperature of the precedent disk. The outside diameter, OD, and thickness of every disk are measured using capacitance gauges, MicroSense, with 10 nanometer resolution. Starting position of the tool on X-axis is corrected, if the OD comes out of the range of ±0.75μm. The self-cutting of the end face of aluminium vacuum chuck is performed to reset the zero position of Z-axis to this face.

*3.2 New technologies for curved surface cutting*

Though the lathe was basically very accurate, we could not get a satisfactory profile at the small radius profile at edge of beam hole (The radius varies from R0.596mm in input-side end disk to 1.089mm in output-side end disk.) in the first stage of this development. We introduced two countermeasures to correct this phenomenon.

First, we developed a new method to evaluate the cutting edge radius. Namely, we cut an aluminium hemisphere by the very tool in order to print its shape on the ball. Then, the circularity of the periphery of cross section cut by a plane including the rotation axis is measured. The deviation of tool radius is deduced from it. Through once or twice passes of this process, sufficient accuracy of cutter radius within 0.3μm was obtained as practical input data of the tool profile term in CNC. Second, the movement of tool edge is optimised for RDDS1 with tuning the parameters in CNC servo-system.

In addition, we introduced a "feed forward" method for "2b" dimension according to the measurement of the acceleration mode frequencies of six-disk set-ups. This method relies on the highly accurate and repeatable machining.

## 4 STACKING AND BONDING

The stacking process of disks is basically the same as the former DDS3 structure stacking. We refurbished the V-block with the fixing jigs and the bookshelf monitoring two axes autocollimator. The disks were stacked on a ceramics cylinder supported by a SUS cylinder along the V-block. After every stacking of 70 or so disks, Viton straps were placed upon the disks and fastened towards V-block by narrow SUS plates and bolts. Finally, the top of the stacked disks is covered with the ceramics cylinder and SUS cylinder and was axially loaded with 600kg by a coil spring. The straightness was measured before transferred to the bonding process.

As for the pre-bonding, the treating temperature of the furnace was a little higher than 150 ºC and total duration was 48 hours. The purpose of this process is freezing the relative motion between disks during the following processes and assisting the better bonding process in later diffusion bonding stage by making bridges between disks without deformation. In the main bonding, the structure was hanged on a ceramics cylinder suspended by three stud rods in a vacuum furnace. The pressure for the bonding was set on the top of the pillar. The top temperature was about 890 ºC, but the duration above 850 ºC was kept for 4 hours. The total process time is about one day.

## 5 RESULTS OF RDDS FABRICATION

The thickness p and OD measured with MicroSense just after turning are shown in Fig.2 and Fig. 3. Variations of the measured values fall between the tolerance zone. Though radial dimensions 2a and 2b are very critical for controlling the frequency characteristics, we inspect the accuracy of every 2a and 2b by CMM (Zeiss CARAT850) after half a day or more later than cutting stage. This means that the in-line process control depends on the inspection of OD. The correlation of 2a and 2b is shown in Fig.4 and this tells that 2a and 2b have a good correlation within 1μm. Further, about two hundreds of 2b measuring data points are plotted in Fig. 5.

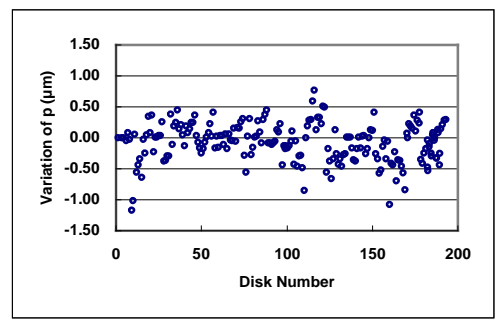

Figure 2: Thickness p variation versus Disk Number

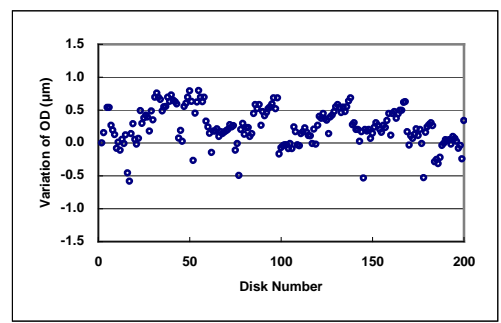

Figure 3: Outside diameter OD variation versus Disk Number

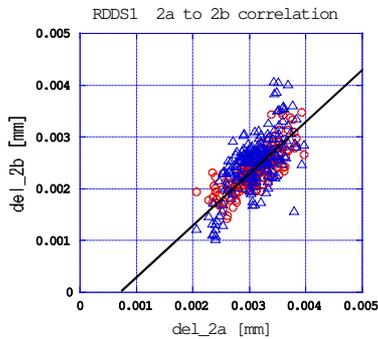

Figure 4: Correlation of 2a and 2b of RDDS1

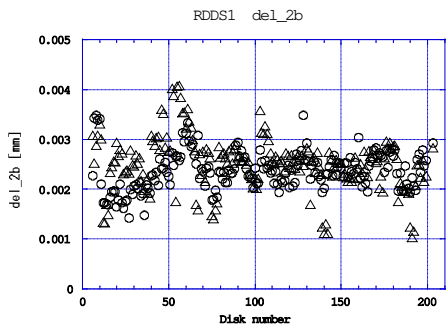

Figure 5: Measuring data of 2b by CMM

The distribution of the data is well within the tolerance of ±2μm and its center value is shifted around 2 to 3μm. This centre value shift is occurred by the overrunning of the measuring probe due to creeping into the annealed soft surface of OFC disk. The diameter of sphere tip of the probe is 1mm and its probing force to the work-piece is 0.2N. After compensation of this shifting, we can see the accuracy of the contour dimension is almost controlled within 2μm. Measuring with interferometer, the flatness was within 0.5μm over 95% of disks. As for the parallelism, 95% of disks were within the tolerance, 17μrad.

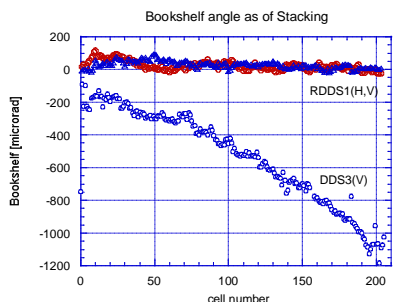

Figure 6: Bookshelving of stacked disks

We achieved a substantial reduction in bookshelving errors during disk stacking, as shown in Fig.6. The cautious stacking operation realized this achievement.

In spite of careful bonding, the bending of the main structure was 200μm almost the same as in DDS3. We observed clearly this time the flare at both ends of the main body as shown in Fig. 7. This is due to the differential thermal expansion of copper and ceramics ($Si_3N_4$) which reacts chemically at a high temperature[3].

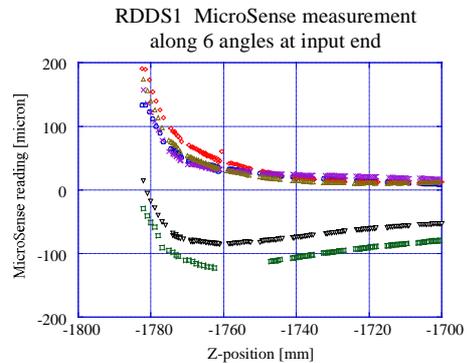

Figure 7: Flare at the main structure input end

## 6 CONCLUSION AND SUMMARY

It was proven that our diamond turning technology could be provided for fabrication of such prototype disks with complicated curved cross section. However we have still a little higher fraction defective on some items and time inefficiency even in the prototype cutting process. The time-consuming stacking process, which is rather labor-intensive, needs to be eventually automated following a precise motion analysis. Issues with flares of the body ends and bowing during the bonding will have to be addressed. In the next stage, we should overcome the difficulties in future mass-production process and cost.

## 7 AKNOWLEDGEMENT

The authors would like to thank Prof. D. Burke of SLAC and Prof. M. Kihara and Prof. K. Kondo of KEK for their continuous encouragement in this fabrication program. We also would like to thank Morikawa Co., Ltd. for their rough turning and fine milling, Ishikawajima Harima Heavy Industries Co., Ltd. for their contribution to stacking and two stages bonding process and other collaborators.